# Chiral and non-chiral swift mode conversion near an exceptional point with adiabaticity engineering

Dong Wang[1,2,3,4,7], Wen-Xi Huang[5,7], Bo Zhou[1,2,3,4], Wenduo Yu[1,2,3,4], Pei-Chao Cao[1,2,3,4], Yu-Gui Peng[5], Zheng-Yang Zhou[6], Hongsheng Chen[1,2,3,4]*, Xue-Feng Zhu[5]*, Ying Li[1,2,3,4]*

[1] *State Key Laboratory of Extreme Photonics and Instrumentation, ZJU-Hangzhou Global Scientific and Technological Innovation Center, Zhejiang University, Hangzhou 310027, China.*

[2] *International Joint Innovation Center, The Electromagnetics Academy at Zhejiang University, Zhejiang University, Haining 314400, China*

[3] *Key Lab. of Advanced Micro/Nano Electronic Devices & Smart Systems of Zhejiang, Jinhua Institute of Zhejiang University, Zhejiang University, Jinhua 321099, China*

[4] *Shaoxing Institute of Zhejiang University, Zhejiang University, Shaoxing 312000, China*

[5] *School of Physics and Innovation Institute, Huazhong University of Science and Technology, Wuhan 430074, China.*

[6] *Key Laboratory of Optical Field Manipulation of Zhejiang Province, Department of Physics, Zhejiang Sci-Tech University, Hangzhou 310018, China*

[7] *These authors contributed equally: Dong Wang, Wen-Xi Huang*

\* e-mail: hansomchen@zju.edu.cn; xfzhu@hust.edu.cn; eleying@zju.edu.cn

**The eigenvalue of a non-Hermitian Hamiltonian often forms a self-intersecting Riemann surface, leading to a unique mode conversion phenomenon when the Hamiltonian evolves along certain loop paths around an exceptional point (EP). However, two fundamental problems exist with the conventional scheme of EP encircling: the speed of mode conversion is restricted by the adiabatic requirement, and the chirality cannot be freely controlled. Here, we introduce a method which dynamically engineers the adiabaticity in the evolution of non-Hermitian Hamiltonians that allows for both chiral and non-chiral mode conversion on the same path. Our method is based on quantifying and controlling the instantaneous adiabaticity, allowing for non-uniform evolution throughout the entire path. We apply our method into the microwave waveguide system and by optimizing the distributed adiabaticity along the evolution loop, we achieve the same quality of mode conversion as conventional quasi-adiabatic evolution in only one-fourth of the time. Our approach provides a comprehensive and universal solution to address the speed and chirality challenges associated with EP encircling. It also facilitates the dynamic manipulation and regulation of non-adiabatic processes, thereby accelerating the operation and allowing for a selection among various mode conversion patterns.**

Non-Hermitian systems can exhibit exceptional points (EPs) where eigenvalues and eigenstates coalesce [1]-[4]. EPs have been extensively studied for their unique properties in the field of photonics[5]-[8], acoustics [9]-[13] and beyond [14], which can be implemented for single-mode lasing [15][16], loss-induced transmission enhancement [17], unidirectional invisibility [18][19]. The eigenvalues of these non-Hermitian Hamiltonians form a self-intersecting Riemann surface, causing the initial eigenstate to be irrecoverable after completing a loop around an EP [20]. Mode conversion can be achieved by adiabatically encircling an EP, but only in one direction (e.g., counterclockwise, CCW) [21]. In the opposite direction (e.g., clockwise, CW), the Hamiltonian transitions to a lower imaginary Riemann sheet, leading to a nonadiabatic jump and subsequent recovery of the initial state [22]-[31]. This asymmetric mode conversion highlights the chirality of EP encircling and has sparked significant research interest, particularly in applications such as optical communications [29], quantum control [30], and optical isolators [31].

Despite notable progress, the scheme of EP encircling faces two fundamental problems. Firstly, the adiabatic requirement limits the speed of mode conversion, impacting operational efficiency and mode decay. Recent approaches using Hamiltonian hopping [32] and selected loops [33] aim to accelerate the evolution process, but they are constrained by extreme parameters and low-loss eigenstates, limiting their general applicability. Secondly, the control over chirality in mode conversion is constrained. Conventional schemes only allow for chiral mode conversion (mode A → B in CCW, mode A → A in CW). Recent findings demonstrate that non-chiral mode recovery (mode A → A in CCW and CW) can be achieved by following a loop path that does not encircle an EP [34], however, the specific outcome depends on the path details. And non-chiral mode conversion (mode A → B in CW and CCW) has also not been adequately observed thus far, as this non-chiral behavior is purportedly achievable only with a specific starting point in the evolution process [35]. Consequently, the speed and chirality challenges in EP encircling have been addressed by a limited number of works that rely on specific models, paths, or Hamiltonians, emphasizing the need for a comprehensive and general solution. Adiabaticity holds the potential to resolve both problems but has not received adequate attention. Existing schemes focus on maintaining proximity to the Riemann surface throughout the encircling operation, with occasional jumps, yet fail to consider the degree of adiabaticity at each instant. A systematic method to quantify and control instantaneous adiabaticity is currently absent, impeding efficiency optimization. Furthermore, non-adiabaticity engineering has been overlooked, with previous approaches either avoiding non-adiabatic processes in mode conversion or limiting non-adiabatic jumps between Riemann sheets in mode recovery. The active introduction and control of non-adiabatic processes remain rare, limiting the ability to accelerate operations and freely select mode conversion patterns. Overall, a comprehensive approach addressing the speed and chirality challenges in EP encircling is urgently needed, with adiabaticity playing a crucial role that requires further attention.

We introduce dynamic adiabaticity engineering for non-Hermitian Hamiltonian evolution, which achieves the same quality as conventional quasi-adiabatic evolution in one-fourth of the time. Our method enables both chiral and non-chiral mode conversion on a single path. By considering the distributed nature of adiabaticity along the parameter path, we optimize the velocity at each local point based on quantified adiabaticity. This approach eliminates the strict adiabatic/non-adiabatic dichotomy, incorporating both processes within the same framework. Importantly, our methodology allows the system state to deviate from the Riemann surface, offering an alternative to strict adherence or jumps between surfaces. Practically, we implement our method in the microwave waveguide system, as an illustration of the applicational insight of our method in the real physical world. It shows the superiority in the speed and mode purity. It is worth mentioning that our approach is not limited to optical systems, it is applicable to any non-Hermitian systems. Our work enhances understanding of adiabaticity in non-Hermitian systems and provides insights for advancing mode converters. The dynamic adiabaticity engineering method is a promising approach for achieving efficient and high-quality evolution in non-Hermitian systems.

*Adiabaticity engineering.* In the study of general chiral phenomena, to complete the switching of the initial mode, the state needs to evolve while adhering to the upper Riemann surface of the imaginary part. This process is typically achieved by evolving at a sufficiently slow speed of parameter change. To quantify this evolving speed, we introduce a function $C(x)$ to represent the position of a N-level system on a certain parameter path: $dC^2(x) = \sum_1^N dx_n^2 /\rho$, $\int_C dC = 1$, where $\rho$ is the normalization factor. Therefore, the speed of parameter change is defined as: $v = dC/dt$. In most research cases, the system moves at a constant speed $v_0$ on the parameter path (the speed is low enough), as illustrated in Fig. 1(a). However, uniform evolution is not the optimal solution to achieve mode conversion because the adiabaticity is not consistent throughout the evolution process. This can be understood by envisioning the parameter path as an arch bridge, where different positions on the bridge have varying "escape velocities", and the magnitude of the "escape velocity" indicates the adiabaticity of that particular position. Adiabatic evolution implies that the system cannot deviate from the bridge surface while traversing the arch bridge.

With this consideration in mind, we can design the nonuniform evolution speed $v(t)$ that depends on the adiabatic requirement along the path: high adiabaticity demands low speed, while low adiabaticity demands high speed. Undoubtedly, this approach can accelerate the mode conversion process (see Fig. 1b). It appears that this adiabaticity-dependent evolution scheme is the fastest method to realize mode conversion if adiabaticity must be preserved throughout the entire evolution path. However, many realistic problems implicate huge demand to rapidly evolve to the target steady state, such as stochastic heat engines [36] and probability distribution of genotypes in a population [37]. In cases where the emphasis is more on the effect of mode conversion, the process can be faster as long as the adiabaticity of a certain evolution path is abandoned. This perspective is vividly illustrated in Fig. 1c. Physically, these three different forms of evolution can be represented on the Riemann surface, as qualitatively depicted in Fig. 1d-Fig. 1e.

Now that different positions along the parameter path have varying adiabaticity requirements, it is necessary to quantify the adiabaticity before investigating the newly proposed dynamic evolution methods. In our study, system's state during evolution is approximately obtained through difference iteration after discretization:

$$|\Psi_j\rangle = \sum_n c_{n,j} |\psi_{n,j}\rangle, \tag{1}$$

$$|\Psi_{j+1}\rangle = \sum_n \langle \theta_{n,j+1}|\Psi_j\rangle e^{-i\omega_{n,j+1}\Delta t_j} |\psi_{n,j+1}\rangle. \tag{2}$$

Here, $\{\langle\theta_n|\}$ and $\{|\psi_n\rangle\}$ are the normalized biorthogonal basis of the non-Hermitian Hamiltonian. The subscript $n$ is sorted from the largest to the smallest according to the imaginary part of the eigenvalue. The subscript $j$ represents the $j$th parameter point, $\Delta t_j$ is the evolving time at the $j$th parameter point, $c_{n,j}$ is the coefficient of eigenstate $\psi_{n,j}$, $\omega_{n,j+1}$ is the corresponding eigenvalue of eigenstate $|\psi_{n,j+1}\rangle$, and $\langle\theta_{n,j+1}|\Psi_j\rangle e^{-i\omega_{n,j+1}dt}$ actually stands for $c_{n,j+1}$. We use the weighted eigenvalue $\overline{\omega}_j$ to characterize the system's state on the Riemann surface:

$$\overline{\omega}_j = \sum_n \frac{|c_{n,j}|^2}{\sum_m |c_{m,j}|^2} \omega_{n,j}. \tag{3}$$

Next, we define the proportion $P_{n,j}$ of the instantaneous eigenstate $|\psi_{n,j}\rangle$ as $P_{n,j} = |c_{n,j}|^2 / \sum_m |c_{m,j}|^2$. Adiabatic evolution generally requires the system to predominantly occupy the least decaying state. In this context, it means that the proportion $P_{1,j}$, corresponding to the set of

instantaneous eigenstates with the largest imaginary parts, should not drop below a specific value throughout the entire evolution time. This value is referred to as the dominant state proportion, denoted as $P_{1,j} \geq P_0$. A higher $P_0$ signifies increased adiabaticity. Varying $P_0$ allows for different evolutionary configurations based on distinct adiabatic requirements. With this definition established, if $P_{1,j} < P_0$, it necessitates an increase in the evolving time $\Delta t_j$ at $j$th parameter point (reduction in evolving speed) to enhance adiabaticity. Conversely, if $P_{1,j} > P_0$, it allows for a reduction in evolving time (increase in evolving speed) or even skipping the parameter point until $P_1$ fall into $P_0$. It is important to note that $P_0$ must be less than 1, as changing system parameters cause coefficients of the instantaneous eigenstates to change. Consequently, the system cannot remain in a single eigenstate without undergoing any changes, rendering complete adiabaticity unattainable.

***Stable conversion.*** With the preset described above, if we set $P_0$ at a constant value and let the dominant state proportion $P_{1,j} = P_0$, we can get the maximum velocity under this adiabaticity requirement, as well as keeping the purity of the end state at a high level. This strategy, which dynamically adjusts the dominant proportion to maintain the purity of the final state, is referred to as the "stable conversion configuration" in this paper.

The residence time $\Delta t_j$ at each parameter point along the evolution path is calculated as follows: Give the current state $|\Psi_j\rangle$, the dominant state proportion at the next parameter point is determined by the equation:

$$P_{1,j+1} = \frac{\left|\langle\theta_{1,j+1}|\Psi_j\rangle e^{-i\omega_{1,j+1}\Delta t_j}\right|^2}{\sum_n\left|\langle\theta_{n,j+1}|\Psi_j\rangle e^{-i\omega_{n,j+1}\Delta t_j}\right|^2}. \tag{4}$$

We then we let $P_{1,j+1}$ equal to $P_0$ to satisfy the adiabatic evolution requirement. This simplifies the equation to a single-variable form with $\Delta t_j$ as the variable. By solving this equation, we can determine the residence time $\Delta t_j$ for each parameter point, and thus obtain the set of evolution times $\{\Delta t_j\}$ (for additional information, see Note 2 in the Supplementary Material).

In this study, we consider a non-Hermitian system with EP:

$$\widehat{H} = \begin{pmatrix} x + iy & 1 \\ 1 & -x - iy \end{pmatrix}. \tag{5}$$

Here $x, y \in \mathbb{R}$. The Hamiltonian of most two-level systems can be transformed into this form (see Note 5 in the Supplementary Material).

The mode conversion results of the stable conversion configuration (nonuniform evolution) are presented in Fig. 2. The evolution time set $\{\Delta t_j\}$ and the evolution trajectory are shown in Fig. 2a and 2b. The parameter path consists of 100 equal intervals, yielding a total of 101 parameter points. Initially, the dominant state proportion $P_{1,j=0} = 1$, reflecting the input state $\psi_{1,j=0}$ and $|\Psi_0\rangle = |\psi_{1,j=0}\rangle$. Consequently, this evolutionary path can be bypassed until $P_{1,j}$ decreases to the value of $P_0$.

In the obtained evolution configuration, we have considered two different input modes and two different evolution directions. These four cases are illustrated in Fig. 2c-2f, where $\zeta_{A,B}$ represents the coefficient of Mode A or Mode B: $\zeta_{A,B;j} = |\langle \theta_{A,B}|\Psi_j\rangle|^2 / \sum_{A,B}|\langle \theta_{A,B}|\Psi_j\rangle|^2$. It is important to note that Mode A and Mode B correspond to the system's two eigenstates on the degenerate axis, namely $|\psi_A\rangle = |\psi_{1,j=0}\rangle$ and $|\psi_B\rangle = |\psi_{2,j=0}\rangle$. Notably, chiral mode conversion effects are observed for both input modes, representing a novel achievement not previously reported. It is important to note that the stable conversion configuration represents an optimization of the conventional method, where the underlying physics responsible for the chiral effect remains rooted in the concept of "nonadiabatic jump" as postulated in conventional chiral conversion. Consequently, the non-chiral effect demonstrated in Section *Swift non-chiral mode conversion* cannot be achieved within this framework.

For comparison, we also present the case of uniform evolution (represented by dashed lines in Fig. 2c-2f). Faster and more pure mode conversion is achieved for Mode A (Fig. 2c). However, the transfer effect is suboptimal for Mode B (Fig. 2e) due to the skipped parameter path at the beginning of counterclockwise (CCW) evolution, which becomes the last path of clockwise (CW) evolution while the evolution time set $\{\Delta t_j\}$ is based on CCW evolution. Additional constraints are required to optimize the effectiveness of CW dynamic evolution.

***Demand-driven strategy for dynamical evolution.*** In dynamic evolution, focusing on adiabaticity in the evolutionary process can effectively maintain the purity of the state. However, strict adiabatic confinement throughout the entire parametric path is often unnecessary. For example, in mode conversion, as long as the final state's purity after switching is sufficiently high, stringent adiabaticity during the path is not crucial (Fig. 1c). This insight has led us to propose a general scheme of dynamic evolution that achieves arbitrary mode conversion effects while ensuring high purity.

Utilizing the stable conversion configuration discussed earlier, two conditions arise during evolution: I. Instantaneous jumps between parameter points, and II. Static evolution at a specific parameter point for $\Delta t_j$. While Procedure I may disregard adiabaticity along certain evolution paths, Procedure II compensates for it. The optimal results of dynamic evolution, tailored to specific requirements, depend on the appropriate combination of jump points and static evolution points using the evolution time set $\{\Delta t_j\}$. Thus, all dynamic evolution problems can be formulated as optimization problems within our framework, differing only in the imposed constraints. In this paper, we summarize the constraints as follows: I. Enclosed parameter path; II. Evolutionary direction (CCW or CW); III. Input mode; IV. Desired end state and its purity. In the realm of optimizing the evolution of quantum states, quantum optimal control theory stands as a highly efficacious approach[38]-[40]. However, quantum control algorithms typically aim to achieve the desired final state without necessarily returning to the original parameter point after system evolution. Given that this constitutes a well-defined optimization problem under fixed closed parameter paths, conventional optimization algorithms are capable of determining the optimal solution. Here, we adopt combined approach utilizing the genetic algorithm [41] and the SQP (Sequential Quadratic Programming) algorithm [42]. For a comprehensive algorithmic description, refer to Note 3 in the Supplementary Material.

*Swift bimodal chiral evolution.* To demonstrate the superiority of our algorithm, we performed bimodal chiral evolution and presented the results in Fig. 3e-3f. The evolution processes of CCW and CW directions in the parameter plane are depicted in Fig. 3a-3b. Notably, the optimized path enables the system to bypass most regions of the evolutionary path and remain at specific parameter points. These parameter jumps occur instantaneously without altering the state in the representation of mode A/B. The mode-switching and mode-recovery phenomena arise from static evolution at these characteristic points, compensating for adiabaticity. By incorporating these jumps and stagnation, the adiabaticity is modulated, resulting in a substantial reduction in evolution time. Comparing the outcomes with uniform evolution, we observe faster mode-switching or mode-recovery effects and higher purity of the end states in all four cases, validating our assumption. It is important to note that these "perfect" results are achieved through the constraints established within our evolution framework (for further details, see Note 4 in the Supplementary Material).

We introduced the chiral index (CI) as a metric to evaluate the quality of the chiral mode conversion effect. The CI is calculated as the average of $\zeta_{A,B}$ for end states in CCW and CW: $CI = \frac{1}{2}\max\{(\zeta_{A;end}, \zeta_{B;end})|CCW\} + \frac{1}{2}\max\{(\zeta_{A;end}, \zeta_{B;end})|CW\}$. A higher CI indicates a stronger chirality in the process. The chiral effects of uniform evolution, stable conversion configuration, and Genetic-SQP Algorithm Optimization for different input modes are illustrated in Fig. 3g. Both the stable conversion and Genetic-SQP methods demonstrate superior chiral effects compared to uniform evolution. Moreover, the Genetic-SQP method offers an advantage over the stable conversion method as it is independent of the input mode, providing greater potential for practical chiral converters.

***Swift non-chiral mode conversion.*** The stable and fast dynamic evolution configuration framework introduced earlier has demonstrated rapid transitions, high purity, and remarkable chiral index in chiral mode conversion effects. It should be pointed out that, our method is a demand-orientated framework designed for dynamic evolution near EP. Beyond chiral evolution, our approach can be extended to achieve various other dynamic evolution effects, including non-chiral evolution, by incorporating corresponding constraints into the optimization process.

By adjusting the imposed constraints (see Note 4 in the Supplementary Material), we achieved non-chiral evolution and obtained faster mode conversion processes, as depicted in Fig. 4. In this context, "chiral" refers to the distinct phenomena evolving in CCW and CW directions, while non-chiral evolution indicates that the output modes are the same for a specific encircling path, irrespective of the evolution direction (CW or CCW).

For Mode A, we successfully realized the mode conversion effect in both CCW (Fig. 4b) and CW evolution (Fig. 4d), which corresponds to the non-chiral mode conversion mentioned in the introduction. Notably, our configuration achieved rapid mode conversion using only a small portion of the parameter path (Fig. 4a). This can be attributed to the significant variation in $\text{Im}(\bar{\omega})$ values along this segment, leading to fast changes in the distribution of instantaneous eigenstates. Therefore, the algorithm concentrated the evolution in this region. The evolutionary behavior is also observed in the imaginary Riemann surfaces (Fig. 4e-4f), indicating the delicate balance between freeing adiabaticity and maintaining the state's purity along the path. We observed fluctuations in $\text{Im}(\bar{\omega})$ on the skipped parameter path even if the system did not evolve in certain parameters, due to the variation in the instantaneous eigenvectors. In contrast, on the parameter path where the system remains, $\text{Im}(\bar{\omega})$ increases.

To highlight the advantages of our approach in term of time consumption, we compare the evolution time required to achieve different final state switching proportions using three methods: Uniform evolution, Stable conversion and Genetic-SQP algorithm optimization under varying final state purity requirements. The results are shown in Fig. 4g. It is evident that the Genetic-SQP algorithm optimization method exhibits significantly reduced time consumption compared to the first two methods. Furthermore, the growth in time consumption is not substantial as the final state purity constraint changes. The advantage of the Genetic-SQP method becomes more pronounced with higher requirements for final purity.

*Microwave waveguide system.* To verify the enormous potential of our strategy in the real physical world, we perform our strategy in a well-known microwave waveguide system as an example [22]. The details and the modulation of this system is presented in the Supplementary Material. Here we reduced the number of constraints and only constrain Mode A convert into Mode B in CCW and CW directions (in waveguide, CCW evolution means the microwave transmit from left to the right and CW evolution, from right to the left). Mode A and Mode B the two eigenmodes of the periodic waveguide at the starting parameter point. The results are shown in Fig. 5. We can observe the dramatic time (length) reduction of our optimized waveguide, compared with the conventional methods in terms of mode conversion (Fig. 5a, Fig. 5c and Fig. 5e). In addition, our method has given a quite right performance in CW evolution, in which the percentage of Mode B is 83% in the optimized waveguide and 61% in the conventional waveguide, indicating the superiority of our method.

In addition to waveguide systems, our method is applicable to various other non-Hermitian systems characterized by non-trivial geometric properties on Riemann surfaces. Furthermore, the optimization objective can be modified accordingly. While in our study, the optimization target is time, in other scenarios such as a quantum heat engine [43], the efficiency (defined as the ratio of net-work output to energy input) could serve as the optimization target function.

In conclusion, we have demonstrated that strict adherence to adiabaticity throughout the parameter loop is not always necessary in the dynamic evolution of non-Hermitian systems, especially when prioritizing mode conversion results or chiral effects. Building upon this insight, we have presented a comprehensive evolutionary framework capable of realizing both chiral and non-chiral phenomena while accommodating any desired purity requirement for the final state. In contrast to traditional encircling approaches, our analytical and computational configuration offers notable advantages, including enhanced speed, model independence, and the ability to generate novel effects. These findings contribute to a deeper understanding of non-Hermitian dynamic evolution. Moreover, the potential application of our method is broad, besides the example demonstration in waveguide system, any N-level systems can make use of this method to optimize computational resources, modulation of intermediate process quantity, and result reconfiguration.

**Acknowledgments**

The work at Zhejiang University was sponsored by the Key Research and Development Program of the Ministry of Science and Technology under Grant 2022YFA1405200, the National Natural Science Foundation of China (NNSFC) under Grants No. 92163123 and 52250191, the Zhejiang Provincial Natural Science Foundation of China under Grant No. LZ24A050002, the Key

Research and Development Program of Zhejiang Province under Grant No.2022C01036, and the Fundamental Research Funds for the Central Universities.19

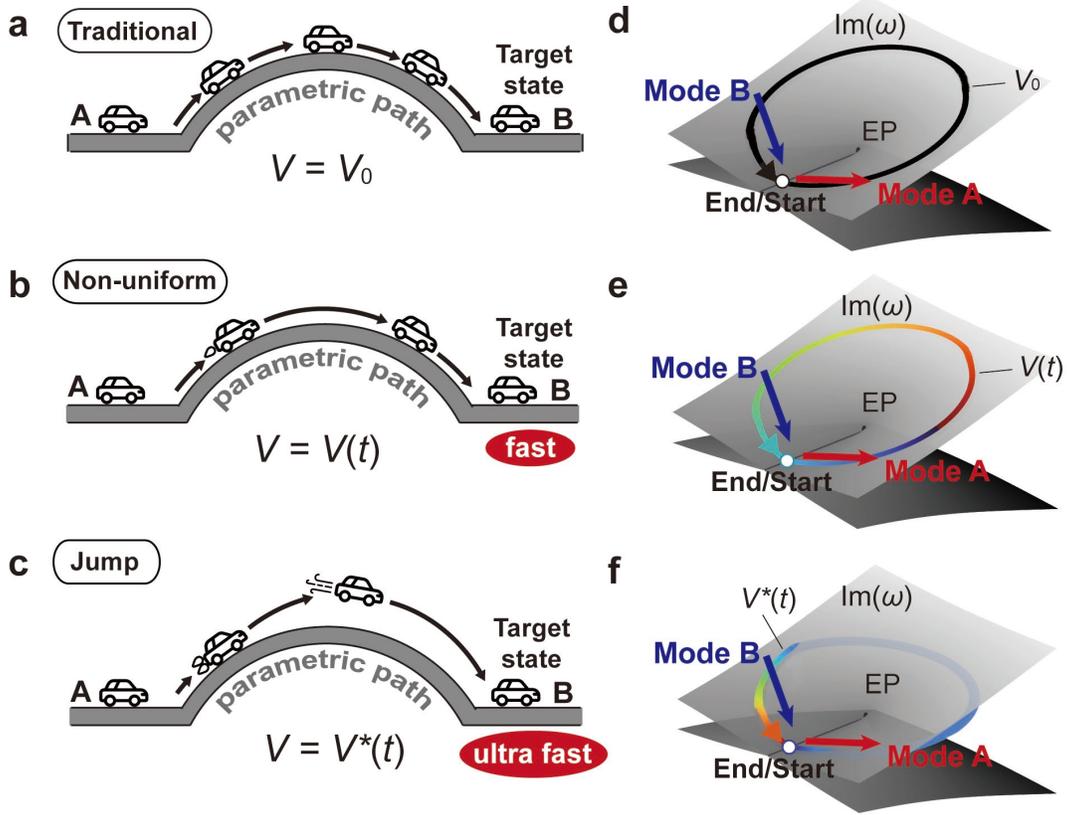

FIG. 1. **The schematic diagram of fast mode conversion through adiabatic engineering along the parameter path**. (**a**) Adiabaticity ensures the system stays on the arch bridge surface but limits its velocity, resulting in slow and uniform evolution. (**b**) Nonuniform speed function $v(t)$ can be designed to consider speed limitations and accelerate the evolution process. (**c**) Adiabatic restrictions can be relaxed at certain positions to achieve faster evolution if the primary objective is reaching the end state. (**d**, **e**, **f**) depict the corresponding cases on the Riemann surface. The system evolves counterclockwise from the branch cut, completes one loop around the EP, and returns to the initial point, transitioning from Mode A to Mode B. The lines represent the system's position in parameter space and depict the variation of evolution speed during mode conversion. (**d**) shows uniform evolution speed $v_0$, (**e**) shows non-uniform evolution speed $v(t)$ determined by adiabaticity at each point, and (**f**) illustrates the "jump" evolution where the system deviates from the Riemann surface at certain parameter points.

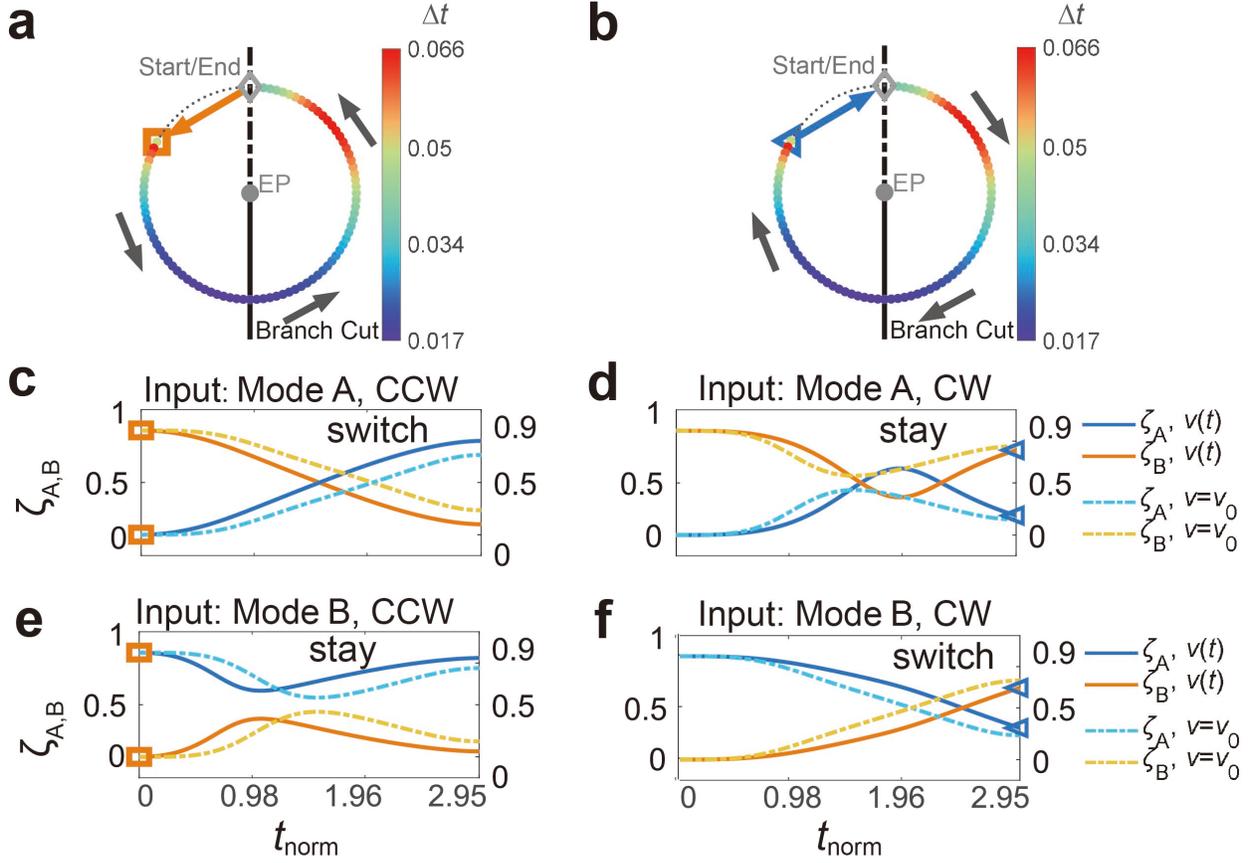

FIG. 2. **Bimodal chiral mode conversion in nonuniform evolution**. (**a**, **b**) show the evolution time set $\{\Delta t_j\}$ for chiral cases and evolution patterns in the parameter planes of CCW and CW directions. In our study, $P_0$ is set to 0.9. The gray rhombus stands for the start/end point. (**c-f**) depict chiral mode conversion effect for input Mode A and Mode B. Mode A switches to Mode B in CCW evolution and remains in Mode A in CW evolution, while Mode B remains in Mode B in CCW evolution and switches to Mode A in CW evolution. The orange box and blue triangle mark the start and end of the actual evolved portion of the system along the parameter path.

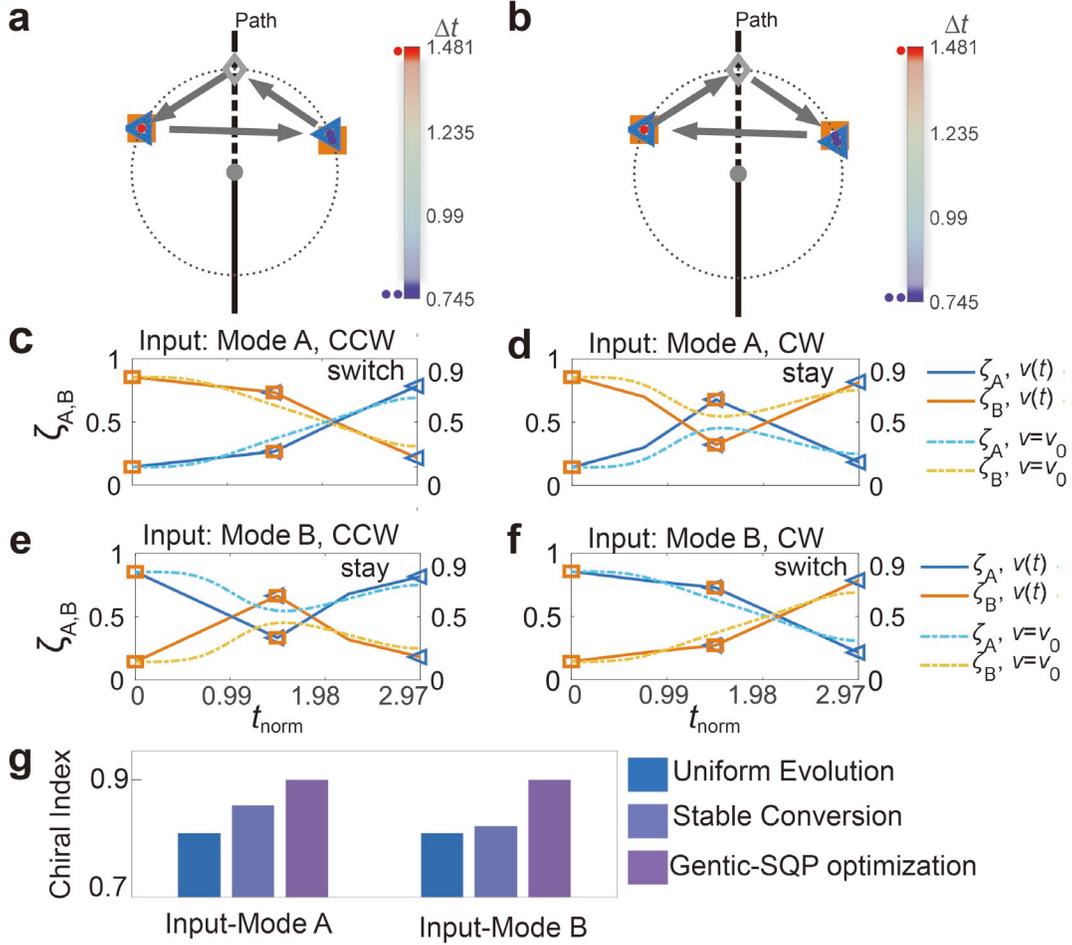

FIG. 3. **Bimodal chiral mode conversion under the optimization of the Genetic-SQP algorithm. (a, b)** display the evolution time set $\{\Delta t_j\}$ for chiral cases and evolution patterns in the parameter planes for CCW and CW directions. The algorithm selects only three parameter points for evolution, skipping a substantial portion of the path (red and blue dots in **a** and **b**). **(c-f)** illustrate the chiral mode conversion effect for input Mode A and B. **(g)** presents a comparison of the chiral effects achieved through uniform evolution, stable conversion, and Genetic-SQP algorithm optimization methods. Our optimized evolution exhibits significant advantages in terms of evolution speed, end state purity, chiral effect, and its robustness (chiral effect being independent of the input mode).

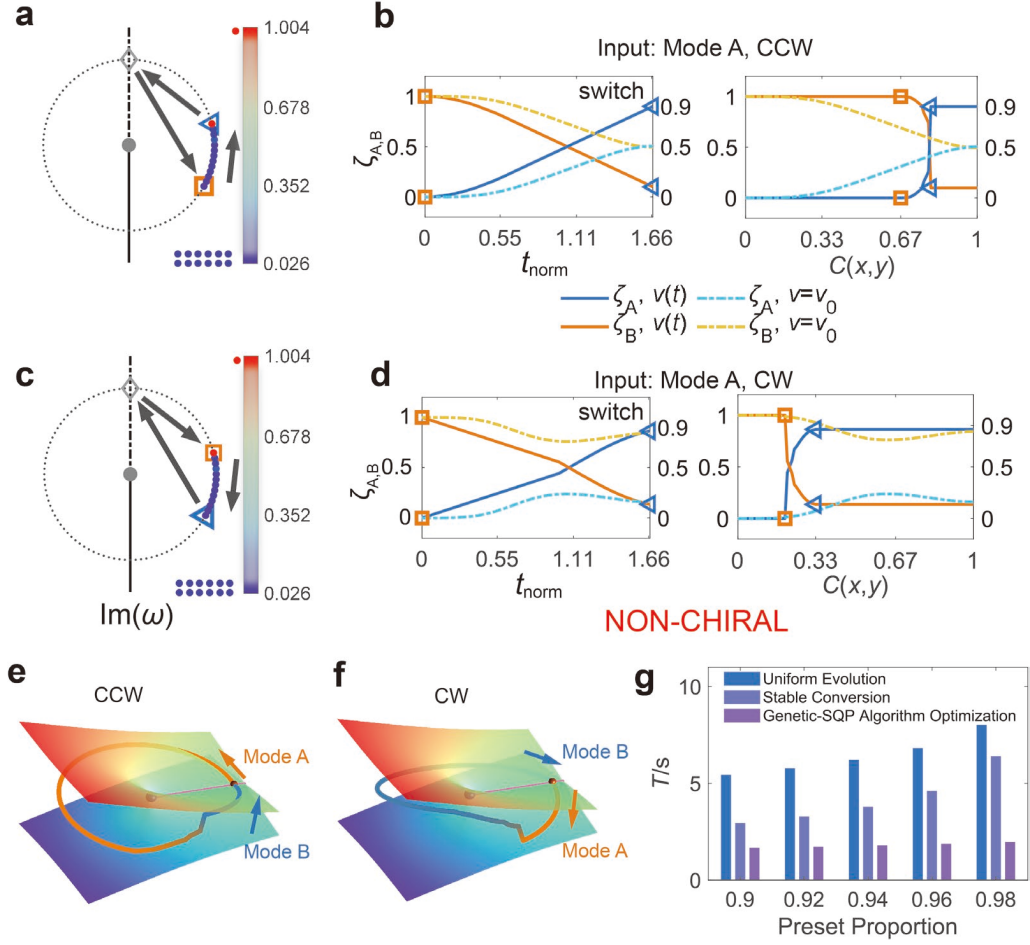

FIG. 4. **Swift non-chiral mode conversion under the optimization of Genetic-SQP algorithm.** **(a, c)** depict the evolution time set $\{\Delta t_j\}$ for non-chiral cases and the evolution patterns in the parameter planes of CCW and CW directions. The algorithm selects a specific portion of the path for evolution (red and blue dots in **a** and **c**) in this case. **(b, d)** showcase the non-chiral state evolution effect for input Mode A. **(e, f)** display the change of the weighted eigenvalue $\mathrm{Im}(\bar{\omega})$ of system's state on the Riemann surfaces for CCW and CW with respect to input mode A. The orange region represents Mode A dominance, while the blue region signifies Mode B dominance. **(g)** compares the time consumption of three evolution methods for achieving mode conversion with different requirements for final state purity.

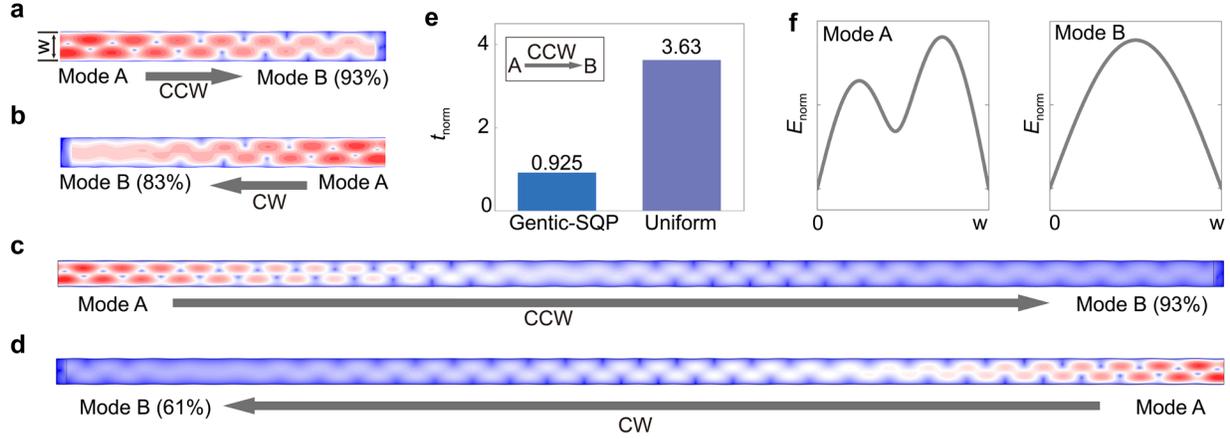

FIG. 5.  **Superiority of the optimization method in microwave waveguides. (a, b)** Optimized modulated waveguide and the corresponding mode conversion results. **(c, d)** Conventional modulated waveguide and the corresponding mode conversion results. **(e)** Time consumption (length of the waveguide) for the optimized and conventional waveguide to realize the mode conversion in CCW parametric evolution. **(f)** normalized electric field of Mode A and Mode B.